\documentstyle[11pt,newpasp,twoside,epsf]{article}
\def\edcomment#1{\iffalse\marginpar{\raggedright\sl#1\/}\else\relax\fi}
\reversemarginpar

\begin{document}
\title{Search for Binary Protostars}
 \author{R. Launhardt$^1$, A.I. Sargent$^1$, H. Zinnecker$^2$}
\affil{$^1$ California Institute of Technology, Astronomy Dept., MS 105-24, Pasadena, CA 91125, USA\\
       $^2$ Astrophysikalische Institut Potsdam, An der Sternwarte 16, D-14482 Potsdam, Germany}


\begin{abstract}
In an effort to shed more light on the formation process of binary stars, 
we have started a program to study multiplicity among nearby low- 
and intermediate-mass protostars using the OVRO Millimeter Array.
Here, we describe the 
project and present the first results on the protostellar core in the Bok globule 
CB\,230 (L\,1177). At 10\arcsec\ resolution, the molecular core is resolved 
into two components separated by $\sim$5000\,AU. The morphology 
and kinematics of the double core 
suggest that it formed from a single cloud core 
due to rotational fragmentation.
\end{abstract}

\section{Introduction}

A major gap in our understanding of low-mass star formation 
concerns the origins of binary stars. 
About 30 to 50\% 
of low-mass main-sequence stars have companions, and the frequency 
of young T\,Tauri binary systems in nearby star-forming 
regions is nearly twice as high.
Binary systems have been observed in all pre-main-sequence 
stages of evolution and there is growing evidence for 
proto-binary systems, although the numbers 
are still very small 
(e.g., Fuller et al. 1996; Looney et al. 1997).
Both theory and observations support the hypothesis that binary 
systems form during the gravitational 
collapse of molecular cloud cores. 
Most scenarios propose bar formation and 
fragmentation in rotating and accreting protostellar cloud cores 
or circumstellar disks as a formation mechanism 
(e.g., Burkert \& Bodenheimer 1996; Boss \& Myhill 1995; Bonnell et al. 1991; 
Boss 1999).
To understand the formation process of binary stars, 
high angular resolution studies of the earliest stages 
of star formation are required. 
We have, therefore, started a program to search for multiplicity among low- 
and intermediate-mass protostars (Class\,0 and I) using the 
Owens Valley Radio Observatory (OVRO) Millimeter Array.

\section{Observations}

Our program  aims at sub-arcsecond resolution corresponding to 
linear resolutions of 150 to 450\,AU. 
Later, with ALMA, we aim
    for 0.1 arcsec resolution, or 15-45 AU, 
    close to the peak of the pre-main sequence
    binary separation distribution. 
The mm continuum emission is used to trace the optically 
thin thermal dust emission.
The molecular gas is traced by the 
C$^{18}$O(1$\rightarrow$0) and 
N$_2$H$^+$(1$\rightarrow$0) lines at 110 and at 93\,GHz, respectively. 
N$_2$H$^+$(1$\rightarrow$0) comprises seven 
hyperfine components and, compared to other molecules, 
depletes later and more slowly onto grains (Bergin \& Langer 1997). 
It is, thus, a very reliable gas 
tracer of the the morphology of protostellar cores.
Initial results presented here are based on observations 
conducted at OVRO in September and October 1999.
The 1\,mm and 3\,mm continuum maps have 1\,$\sigma$\ rms sensitivities of 
4\,mJy/beam for HPBW 2.0\arcsec$\times$1.5\arcsec\ and 
0.7\,mJy/beam for HPBW 5.2\arcsec$\times$4.2\arcsec, 
respectively.
The N$_2$H$^+$\ images were obtained at low resolution only and 
have a velocity resolution of 0.2\,km/s and a 1\,$\sigma$\ sensitivity of 
110\,mJy/beam for HPBW 13\arcsec$\times$9.4\arcsec.

The NIR, submm, and 1.2\,mm continuum observations were performed 
at the 3.5\,m Calar Alto telescope (MAGIC), the 15\,m JCMT (SCUBA), and 
the IRAM 30\,m telescope (19-channel bolometer array). 
We wish to thank Th. Henning, R. Zylka, R. Lenzen, D. Ward-Thompson, 
and J. Kirk who are involved in these programs.

\section{CB\,230 (L\,1177)}

CB\,230 
is a Bok globule located at a distance of $\sim$450\,pc. 
It contains a strong submm/mm continuum source 
(Launhardt \& Henning 1997; Launhardt et al. 1998, 2000) 
and a dense CS core which shows spectroscopic 
signature of mass infall (Launhardt et al. 1997). 
The dense core is associated with two NIR reflection 
nebulae separated by $\sim$10\arcsec\ 
(Yun 1996; Launhardt 1996). 
The western nebula is bipolar with a bright northern lobe 
perfectly aligned with the blue lobe of 
a well-collimated CO outflow (Yun \& Clemens 1994). 
The much fainter southern (red) part of this bipolar 
nebula seems heavily obscured possibly by the infalling envelope.
No star is visible and  
the NIR morphology can be interpreted as light emerging from a 
deeply embedded YSO and scattered outward through the outflow cone 
directed towards us. 
The eastern NIR nebula is much fainter and redder 
and displays no bipolar structure. 

\begin{figure} 
\plotone{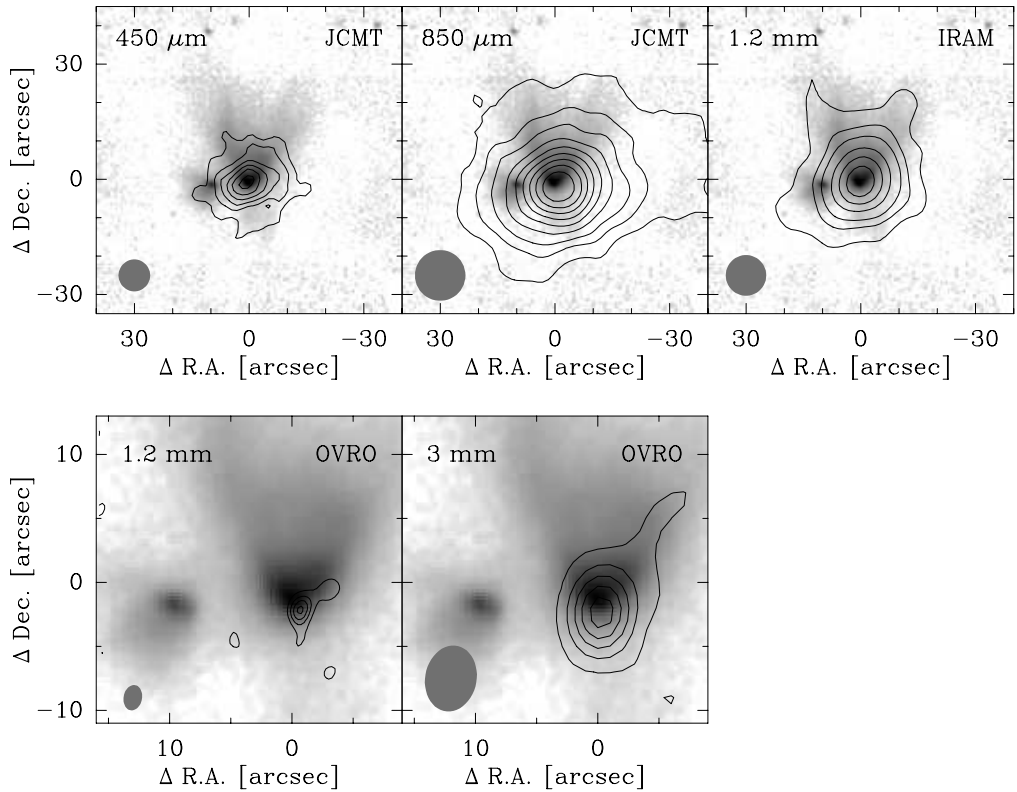}
\caption{\small K-band image of the NIR reflection nebula in CB\,230 (grey-scale, 
Calar Alto 3.5\,m telescope, MAGIC) 
overlayed with contour maps of the dust continuum emission at 
450 and 850\,$\mu$m (JCMT), 
1.2\,mm (IRAM; OVRO), and at 
3\,mm (OVRO).
Contour levels start at 3\,$\sigma$.}
\end{figure}

Previous single-dish mm continuum and molecular line observations did not resolve 
the central part of the dense core. But they demonstrated that the mm emission 
has a core-envelope structure and peaks at the origin the western bipolar NIR 
nebula (Fig. 1, top row). 
The slight extension of the continuum emission to the south east 
suggests that the faint eastern NIR source is also associated with 
circumstellar material. 

The new OVRO continuum maps at 1\,mm and 3\,mm (Fig. 1, bottom row) 
show only one unresolved component clearly associated with the origin 
of the western bipolar nebula. The compactness and location of the 1\,mm continuum source
observed ($<$\,400\,AU E-W extent), together with the bipolar structure 
of the NIR nebula, suggest the presence of a circumstellar disk. The compact source contains 
$\sim$10\% of the total 1\,mm continuum flux in the IRAM map. A significant contribution 
by free-free emission can be ruled out since the bolometric luminosity of the entire 
cloud core of 11\,L$_{\odot}$\ points to a low-mass protostar 
with no capability to ionize its environment (Launhardt et al. 1997). The eastern source 
may be too faint to detect ($< 2$\,mJy at 3\,mm and $< 10$\,mJy at 1\,mm) 
or no compact disk is associated with it. 

\begin{figure} 
\plotone{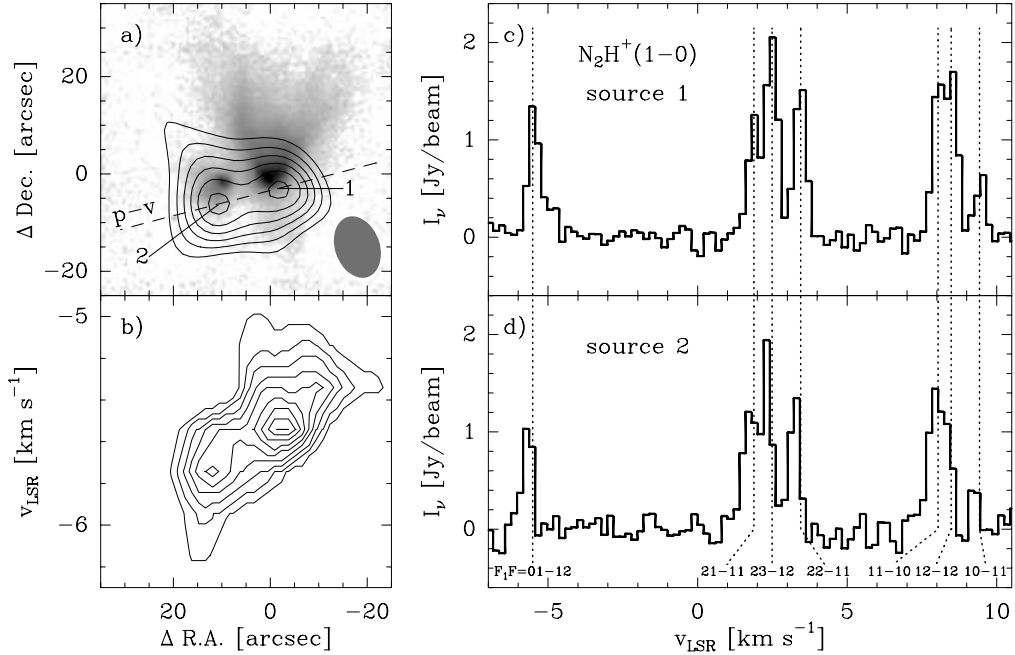}
\caption{CB\,230, N$_2$H$^+$(1--0) OVRO results. 
a) Contour map of the integrated intensity of the 
(J\,F$_1$\,F\,=\,101--012) hyperfine structure line of N$_2$H$^+$\ together with 
K-band image of the NIR reflection nebula (grey-scale).
b) Position-velocity diagram of this line component 
along the dashed line marked in box a). 
c) and d) Full spectra of the N$_2$H$^+$(J=1--0) line 
at the two source positions marked in map a).}
\end{figure}

In contrast to the dust continuum emission, all seven hyperfine structure 
components of the N$_2$H$^+$(1--0) line are detected at both NIR positions. 
The N$_2$H$^+$\ data resolve the molecular cloud 
core into two separate components each of which is spatially coincident 
with one of the two NIR nebulae (Fig. 2). 
The projected separation of the two sources is $\sim$5000\,AU. 
The double core seems to rotate around an axis perpendicular to the connecting line 
and approximately parallel (in projection) to the outflow of the western source. 
A comparison of the kinetic, gravitational, and rotational energy of the double core 
system shows that the two cores are gravitationally bound. 
This is consistent with the assumption that the double core formed due to 
rotational fragmentation from a single cloud core and with the orientation 
of the assumed circumstellar accretion disk around the western protostar.
The angular resolution is not yet high enough to 
derive the rotation curves of the individual cores, but planned observations 
should improve the resolution considerably.
The projected separation of $\sim$5000\,AU is at  
the upper end of the pre-main-sequence binary separation distribution, 
Nevertheless, these preliminary results suggest that the Bok globule CB\,230 contains a ``true'' 
wide binary protostar system.

\end{document}